%% file: workshop_paper_2021.tex
\documentclass[conference,10pt]{IEEEtran}
\IEEEoverridecommandlockouts

\usepackage{amsmath,amssymb,amsfonts}
\usepackage{multicol}
\usepackage{cite}
\usepackage{algorithmic}
\usepackage{graphicx}
\usepackage{textcomp}
\usepackage{xcolor}
\usepackage{hyperref}
\usepackage{xcolor}
\usepackage[all]{nowidow}

\newcommand{\mytextapprox}{\raisebox{0.5ex}{\texttildelow}}

\graphicspath{ {./images/} }

\def\BibTeX{{\rm B\kern-.05em{\sc i\kern-.025em b}\kern-.08em
    T\kern-.1667em\lower.7ex\hbox{E}\kern-.125emX}}
\begin{document}

\title{On Generating and Labeling Network Traffic with Realistic, Self-Propagating Malware
\thanks{The research reported in this document/presentation was performed in connection with contract number W911NF-18-C-
0019 with the U.S. Army Contracting Command - Aberdeen Proving Ground (ACC-APG) and the Defense Advanced Research Projects Agency (DARPA).}
}

\author{
  \IEEEauthorblockN{
     Molly Buchanan, 
     Jeffrey W. Collyer, 
     Jack W. \\ Davidson, 
     Jason D. Hiser, 
     Alastair Nottingham 
  }
  \IEEEauthorblockA{
\textit{University of Virginia},
    Charlottesville, VA, U.S.A.  \\
    \{mkb4vb,jwc3f,jwd,hiser,atn5vs\}@virginia.edu
  }
  \and
  \IEEEauthorblockN{
     Saikat Dey, 
     Mark Gardner, 
     Jeffry Lang 
  }
  \IEEEauthorblockA{
    \textit{Virginia Tech},
    Blacksburg, VA, U.S.A.  \\
    \{dsaikat,mkg,jefflang\}@vt.edu
  }
  \and
  \IEEEauthorblockN{
     Alina Oprea
  }
  \IEEEauthorblockA{
    \textit{Northeastern University}, \\
    Boston, MA, U.S.A. \\
    a.oprea@northeastern.edu
  } 
  
}

\maketitle

\pagenumbering{arabic}
\thispagestyle{empty}
\pagestyle{plain}

%
\input{abstract}


\input{introduction}
\input{data_collection}

\input{attack_recreation}

\input{labels}
\input{related}

\input{summary}

\bibliographystyle{./bibliography/IEEEtran}
\bibliography{./bibliography/IEEEabrv,./bibliography/IEEEexample,./bibliography/atn}

\end{document}

%% file: abstract.tex
\begin{abstract}
Research and development of techniques which detect or remediate malicious network activity require access to diverse, realistic, contemporary data sets containing labeled malicious connections. In the absence of such data, said techniques cannot be meaningfully trained, tested, and evaluated. Synthetically produced data containing fabricated or merged network traffic is of limited value as it is easily distinguishable from real traffic by even simple machine learning (ML) algorithms. Real network data is preferable, but while ubiquitous is broadly both sensitive and lacking in ground truth labels, limiting its utility for ML research.

This paper presents a multi-faceted approach to generating a data set of labeled malicious connections embedded within anonymized network traffic collected from large production networks. Real-world malware is defanged and introduced to simulated, secured nodes within those networks to generate realistic traffic while maintaining sufficient isolation to protect real data and infrastructure. Network sensor data, including this embedded malware traffic, is collected at a network edge and anonymized for research use.

Network traffic was collected and produced in accordance with the aforementioned methods at two major educational institutions.  The result is a highly realistic, long term, multi-institution data set with embedded data labels spanning over 1.5 trillion connections and over a petabyte of sensor log data.  The usability of this data set is demonstrated by its utility to our artificial intelligence and machine learning (AI/ML) research program.

\end{abstract}

%% file: introduction.tex
\section{Introduction}

Computing networks face a variety of attacks, but large network traffic volumes and the need to keep connected devices operational make it infeasible for limited personnel to manually detect and respond to every attack.  Though desirable, automatic or semi-automatic attack detection and remediation techniques are difficult to develop due to the dearth of publicly available, representative data sets \cite{lack-of-data}.  Considering their ability to model network traffic and pick out unusual patterns, AI/ML techniques are particularly suited to automatic network attack detection.  AI methods can be \emph{supervised} (learning to classify malicious and benign traffic)~\cite{exposure,made}, or \emph{unsupervised} (learning the normal traffic distribution and detecting anomalies)~\cite{beehive,portfiler}. Supervised AI techniques rely on the availability of ground truth samples of malicious traffic for training; both methods need malicious samples for testing and evaluation. However, infected machines usually generate copious legitimate traffic, thus necessitating fine-grained labeling of the exclusively malicious activity.   One of the main challenges in developing AI/ML for network security is \emph{obtaining realistic network traffic including benign and malicious connections with fine-grained labeling}~\cite{sommerpaxson}. 

Of course, one cannot simply permit unmodified malware to infect in-use machines and record the ensuing traffic to generate the necessary data.  A real-world network cannot reasonably and responsibly risk the operational interruption and data loss that malware can cause.  But continuously recording traffic and waiting for an inevitable infection is also insufficient due to the uncertainty and difficulty of correctly identifying malicious traffic.  Modern tools like VirusTotal are not adequately accurate to provide sufficient information to determine which network nodes were infected and when, or which traffic from said nodes was malicious~\cite{virustotal}.

To address the AI/ML requirements and concerns of using real malware on a real network, these data sets can be generated by introducing defanged malware into the real network and capturing and anonymizing the resulting traffic, which will include a realistic mix of benign and malicious traffic. The remainder of the paper describes how this goal was achieved to facilitate AI/ML research and experimentation.  

\noindent The major contributions of this work are:
\begin{itemize}
    \item a workable system to capture and anonymize network traffic,
    \item a model to safely deploy malware in a real network, including a realization of the Mirai virus, and
    \item a model for fine-grained labeling of the generated malicious traffic in the network capture.
\end{itemize}

%% file: data_collection.tex
\section{Anonymous Data Collection and Use}

This section briefly explores the collection and anonymization of enterprise scale network sensor data to produce a diverse, realistic, long-term data set for developing and training machine learning algorithms and AI approaches. 

\begin{figure}[bhtp]
\small
\centering
\includegraphics[trim=0in 0.25in 0in 0.1in, clip,width=\columnwidth,height=3.5cm]{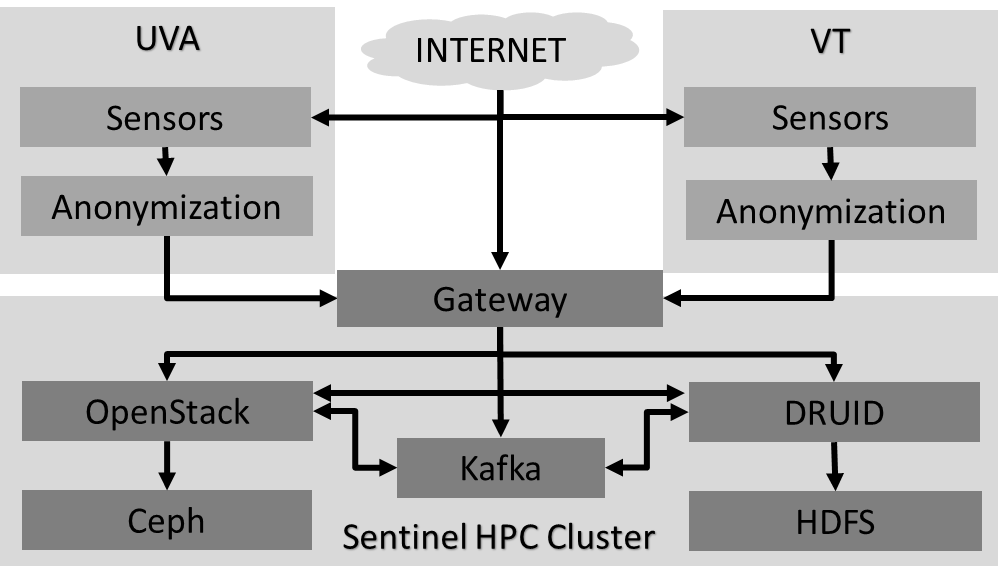}
\label{fig:data-collection-process}
\caption{Overview of the collection, anonymization and processing of network sensor data.\label{fig:data-collection-atn}}
\vspace{-.2in}
\end{figure}

The data set is composed of anonymized network sensor data from two large, disjoint, and geographically separated enterprise networks at the University of Virginia (UVA) and Virginia Tech (VT); these networks differ in topology, service and application ecosystems, and baseline behavior. Figure \ref{fig:data-collection-atn} provides an overview of the data collection and anonymization process.


\subsection{Data Collection}

The generated data set includes a variety of network sensor and appliance logs that collectively provide a detailed trace of the network state and user/device behavior. While packet traces may provide a more complete view of network behavior and additional metrics, sensor logs are preferred because they are (a) pre-aggregated and normalized, which simplifies anonymization without sacrificing application level detail, (b) appreciably smaller and more compressible, making long term storage less expensive, and (c) significantly less computationally expensive and complicated to process at enterprise scale.  

Zeek \cite{zeek-web} provides the primary source of network activity data; it generates a wide selection of inter-connected logs across a broad range of protocols and field types. Zeek logs are supplemented by a range of other sensor and appliance logs that provide context regarding the network's state. These include logs generated by FireEye \cite{fireeye} --- which enumerate detected threats such as known malware callbacks --- and STINGAR honeypot logs \cite{stingar} --- which provide community-sourced threat intelligence from a distributed honeypot network. Together, these provide a valuable source of ground truth to seed discovery of natural malicious behavior in Zeek logs. UVA additionally collects NAT \cite{RFC3022} (Network Address Translation), DHCP (Dynamic Host Configuration Protocol), and asset registration logs to map between public and private addresses.

The creation of this data set is achieved through partnership with and oversight from Information Security departments and professionals at UVA and VT. As network traffic transitions across the UVA and VT network edges, traffic is mirrored through a packet broker to sensor appliances within the respective networks for log creation. All logs are archived, allowing future retrieval and reprocessing when necessary.  

Though collection is broadly similar at both institutions, it differs in traffic composition, the types and placement of sensor appliances, and how data is transitioned through anonymization to be made accessible to researchers. Some of the primary differences between UVA and VT data collection are shown in Table \ref{tab:uva-vs-vt}.


\begin{table}[bhtp]
\centering
 \begin{tabular}{|c | c c|} 
 \hline
 & UVA & VT \\ 
 \hline
 \multicolumn{3}{|c|}{Network Border Traffic} \\
 \hline
 IPv6 & No & Yes \\  
 NAT Type\cite{RFC3022} & Carrier Grade NAT & Traditional NAT \\
 \hline
 \multicolumn{3}{|c|}{Sensor Logs} \\
 \hline
 Zeek & Yes [Corelight \cite{corelight}] & Yes [Custom]  \\ 
 FireEye & Yes & No \\
 STINGAR & Yes & Yes \\
 NAT, DHCP, Asset & Yes & No \\
 \hline
 \multicolumn{3}{|c|}{Current Zeek Dataset (Combined)} \\
 \hline
 Connections Records& \multicolumn{2}{c|}{\mytextapprox 1.5 Trillion} \\
 Total Volume &  \multicolumn{2}{c|}{\mytextapprox 1 Petabyte} \\
 \hline
 \end{tabular}
 \caption{Summary of UVA and VT data collection.}
 \label{tab:uva-vs-vt}
\vspace{-.4in}
\end{table}

\subsection{Log Anonymization}

Collected network sensor data contains significant PII (Personally Identifiable Information) and other sensitive information. To protect network users and organizational security, sensitive fields must be obfuscated or removed from logs before the data set can be made accessible to researchers. Anonymization through overwriting or otherwise destroying field data \cite{dag-scrubber, scrub-tcpdump}, while effective, significantly reduces the fidelity of anonymized records \cite{graded-forensics}. In particular, fields anonymized by destructive methods cannot be leveraged effectively in inter-record and cross-network comparisons. 

To retain a measure of fidelity and comparability in field data, the removal or destruction of field values is avoided wherever possible. Instead, a combination of careful field parsing and one-way cryptographic functions are used to eliminate identifying data while preserving many of the features and patterns inherent in the raw logs. Such functions include prefix preserving IP address anonymization \cite{cryptopan, cryptopant} and custom hash-based functions for other field types. Hashes use a 256-bit secret key as salt to (a) add entropy to fields and protect against collision-based attacks on anonymization and (b) ensure consistency across log types and institutions over time. Anonymization is subject to extensive testing to identify emergent issues, and is regularly updated and improved. When necessary, archived logs are re-anonymized to take advantage of new features and correct identified anonymization errors. 

While every effort is made to ensure all data elements are correctly anonymized, the complex nature of the produced data set presents an unavoidable known risk of anonymization errors in edge cases. This risk is addressed through mitigating policy and infrastructure-based controls to protect edge cases when anonymization is insufficient or otherwise fails. 

\begin{figure}[bhtp]
\centering
\includegraphics[trim=0 0in 0 0 ,clip,width=\columnwidth]{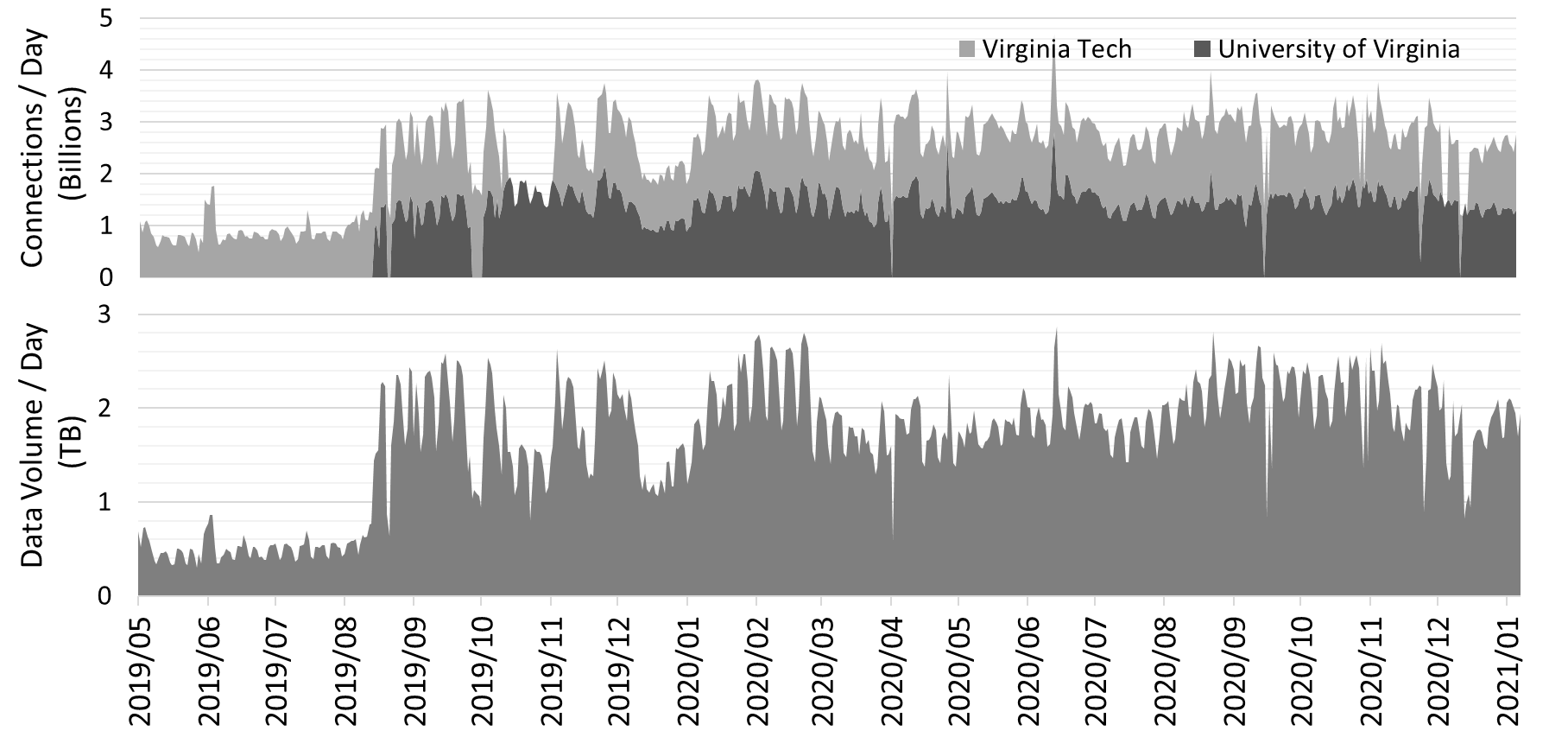}

\caption{Data volume and number of distinct connections per day for the combined Zeek data set. It currently takes between 30 and 45 minutes to anonymize a days worth of traffic at a given institution using a single server. \label{fig:data-summary}}
\vspace{-.25in}
\end{figure}

To this end, sensor data is only made available on Sentinel: a dedicated, access-controlled High Performance Computing (HPC) cluster provisioned to support AI/ML research (see Figure \ref{fig:data-collection-atn}). Data access is limited to approved researchers and subject to restrictions that prohibit data exfiltration and govern how records may be used in publications. Anonymized sensor data is published daily (i) as compressed files, (ii) in Apache Kafka \cite{kafka} as a message streams, and (iii) in Apache Druid \cite{druid} which supports fast SQL-like (Structured Query Language-like) queries. Figure \ref{fig:data-summary} provides a high-level view of the data set available on Sentinel. 

As the data collected at the network edge is composed of real traffic from institutions that actively defend their networks, actual instances of known malware campaigns and malicious traffic are uncommon. The following sections present a method for safely and realistically introducing attack traffic into sensor logs through attack recreation exercises. These provide additional and verifiable ground truth for malicious activity in the data set.

%% file: attack_recreation.tex
\section{Attack Recreation}
A key aspect of the proposed data set production approach is the introduction of realistic malware traffic to the network.  Said generated traffic should combine with the network's natural traffic using standard network protocols, yet machines and appliances should be secure from compromise.  To achieve this balance of realism and safety, real-world malware is defanged and carefully deployed on real networked systems.

This approach has two key advantages.  First, as malware traffic is naturally mixed in the network, no artifacts from mixing are evident in the data set.  This feature is important because otherwise machine learners can easily determine which traffic is which by examining artifacts unintentionally added during the traffic merging process.  Second, the malware never resides on operational systems with sensitive data.  This feature is essential to obtain permission for deployment from network operators, as normal network operation can not be interrupted, and intentional data exposure will not be tolerated.

\begin{figure}[bhtp]
\centering
\includegraphics[width=0.9\linewidth,trim=2.75in 2.25in 1.75in 2in,clip]{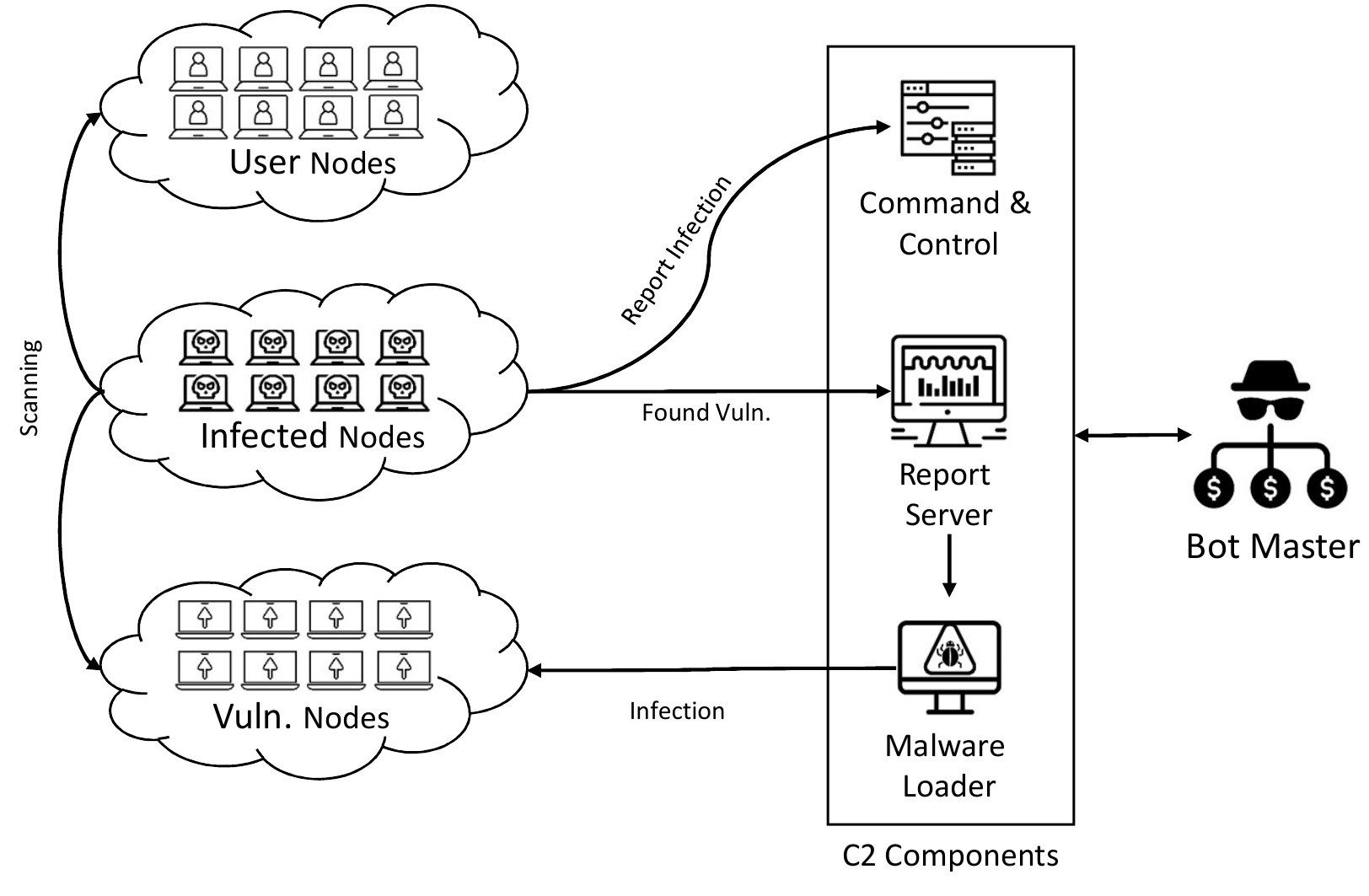}
\caption{Mirai malware, as deployed for collecting datasets\label{fig:mirai}}
\vspace{-.25in}
\end{figure}

As a proof of concept, the Mirai virus was defanged and deployed on the University of Virginia (UVA) and Virginia Tech (VT) networks.  Mirai, as depicted in Figure~\ref{fig:mirai}, operates by leveraging already infected nodes to scan the network to locate new nodes that respond to telnet or ssh.  It then brute-forces the passwords using a built-in dictionary of common/default credentials~\cite{kolias2017ddos,antonakakis2017understanding}.  When a new vulnerable node is detected, an infected node reports this to the report server, which instructs the malware loader to infect the vulnerable node.  After infection, the vulnerable node reports to the command and control server as a bot; at this point, the bot master can leverage the new addition to its botnet for other purposes, such as a DDoS attack.

Mirai is an excellent candidate for experimentation because it was particularly successful real-world malware, variants of which continue to be seen in the wild.  Mirai infected over 65,000 on its first day of deployment and totaled over 600,000 nodes at its peak. Additionally, since Mirai and its control components were open-sourced, it was easy to safely defang.
Mirai was defanged by:
1) Removing portions of the infection that might cause damage.
2) Replacing the default credential dictionary with 20 and 25 character randomly-selected alpha-numeric usernames and passwords, respectively, which were statistically unlikely to match any real-world accounts. 
3) Adjusting the malware loader to safeguard user nodes by only permitting infection of nodes in the pre-selected vulnerable nodes pool.
4) Limiting scanning traffic to permissible and non-critical nodes - for example, UVA's hospital system is not scanned.

To provide further evidence to the UVA and VT network and security groups, unit and integration tests were developed for defanged components. Test deployments in a sandbox verified that only acceptable IPs would be contacted.  While these techniques are specific to Mirai, safely handling other malware should be possible, especially via careful application of network firewall rules.

The precautions installed in Mirai were sufficient to convince UVA and VT's network operation centers to allow Mirai to scan a large subset of public IP space.  The defanged Mirai has scanned a large swath of over 500,000 IPs at UVA and VT.

Nodes sending only malicious traffic might be conspicuous to machine learners, so the Caldera \emph{human} plugin was run to add mitigating benign traffic.  The plugin randomly browses websites, performs google searches of random sentences, and provides realistic delays between operations~\cite{caldera:human,booz2020towards,mitre:caldera}. 

To mimic an outside attacker's command and control (C2) server, C2 components were positioned on the third-party NSF-funded Chameleon Cloud system~\cite{keahey2020lessons}.  Traffic generated as infected nodes in each network contact external C2 components or scan vulnerable nodes in the adjacent network is collected as it transits the network edge, resulting in realistic merging of the attack traffic in sensor logs.

%% file: labels.tex
\section{Generated Labeled Traffic}

A key requirement for both training and evaluating AI/ML algorithms is labeled data.  Each connection captured in the anonymized logs includes associated metadata which determines whether it was natural traffic on the network or generated during the attack recreation.  Ideally, separate labels would be generated for each phase of the attack; for example, scanning traffic might be labeled differently than infection traffic that downloads malware to a new system.

In this experimental setup, such labels can typically be applied in a post-hoc manner by inspecting particular fields in the network traffic and applying a set of simple heuristics.  If Mirai uses port 80 of a particular web server to download the malware from a particular set of URLs, simply filter traffic that matches those criteria and label them as malware download traffic.

A library was developed to label traffic captured during a Mirai attack recreation event according to the aforementioned filtering technique.  For the defanged Mirai, scanning traffic occurs on ports 23 and 2323, virus download happens from a web sever set up on port 80, and C2 communication happens on ports 23 and 48101.  Furthermore, a record is kept of the (anonymized) IPs of all the vulnerable nodes in the experiment.  Any traffic not to or from these IPs is labeled as \emph{natural}.  Traffic to/from these IPs which matches one of the above rules is labeled as indicated; other traffic from the experimental nodes is labeled as generated, benign traffic.

These simple rules were able to label 99.9\% of all traffic during a 6 hour experiment.  Only a handful of unlabeled connections remained, all of which seemed to be oddities of the capture process (e.g., when traffic capture was started during an already active connection).

One might wonder whether the network's natural traffic is benign or could contain traffic from an infected node.  In a network of sufficient size and complexity, it is likely that some traffic is malicious (hence the application of the natural rather than benign label.)  However, the traffic labeled malicious during attack recreations is verifiable so, and can be accurately used for training and evaluation of AI/ML techniques.  Further, during evaluation of a (good) algorithm, little traffic should be detected as malicious; such traffic can (and should) be analyzed by an expert to determine whether it is truly malicious or a false positive detection.  In fact, this type of event happened during the associated AI/ML research, where previously unknown malware was detected in the network.

The labeled Mirai attack was used to evaluate an ML-based system for self-propagating malware detection~\cite{portfiler}. The PORTFILER system creates aggregated network traffic features for a set of monitored ports, learns the legitimate network profiles on those ports during training in an unsupervised manner, and detects and ranks anomalies at testing time. Labeling the different attack stages (e.g., scanning, C2) was particularly useful for evaluating this detector. In use, PORTFILER  detected scanning traffic with high precision and recall and low false positive rates; detection results were more accurate at higher scanning rates. The detailed analysis of PORTFILER's performance was made possible by fine-grained labeling of individual connections and attack stages. Moreover, by analyzing the traffic of all infected machines, it was possible to identify and confirm the C2 server IP based on the number of connections on port 23 and the large duration of those connections compared to legitimate connections. In future experiments, likewise defanged evasive variants of Mirai and other malware will be generated, and novel ML techniques will be developed to detect them more robustly.

%% file: related.tex
\section{Related Work}

While  IP address anonymization is well understood \cite{cryptopan, cryptopant, ontas} and significant related work exists in the area of network traffic anonymization, most approaches focus on packet traces. A wide variety of approaches to trace anonymization have been developed, both in software \cite{pcap-lib, tcp-anon, scrub-tcpdump, tcpkpub, caida-anon-tools, dag-scrubber, lander, generic-anon, bro-anon}, and in hardware \cite{ontas, transparent-anon, netfpga-anon}, although many of these approaches are now defunct. While some implementations focus on appliance and service logs \cite{netflow, canine-1, canine-2, flaim, dnsanon}, these approaches are typically quite destructive as they are intended for broad sharing. Published available data sets \cite{lanl-unified, lanl-comprehensive, unsw} tend to have a limited duration, are more specialized and/or synthetic, and are not representative of normal enterprise traffic \cite{lack-of-data}. 

Previous work on generating and labeling data synthetically also exists~\cite{Landauer2020HaveIY,Sharafaldin2018TowardGA,Choi2020ExpansionOI,Haider2017GeneratingRI,Skopik2014SemisyntheticDS}.  Synthetic techniques suffer to various degrees with artificial artifacts, but issues with unrealistic data have been generally accepted in light of the working assumption that real data on real networks is too cumbersome to generate and too sensitive to provide to researchers.  The approach outlined in this paper contrasts said assumption by demonstrating the viability of a more natural means of data production and illustrating its successful use in ML research.

%% file: summary.tex
\section{Future Work and Summary}

A good way to create data sets for AI/ML training and evaluation is to outfit existing networks for collection and anonymization. This approach can be achieved with commodity hardware and software solutions. Malicious traffic can be carefully added to the existing network by defanging real malware and controlling where in the network it is deployed.  Labeling the traffic can be achieved through simple heuristics based on knowledge of the malware's behavior and the known IPs of the infected machines.

Utilizing this approach and working with UVA and VT's network operators, extant networks were instrumented, traffic collection and anonymization were measured, defanged Mirai was deployed, and secure access infrastructure for researchers was provided.  Simple labeling heuristics were able to generate useful labels for internal research programs; their success in using the produced data set to train models that detected malware demonstrates the proposed approach's feasibility.

In the future, the range of collected sensor data will be expanded, and field coverage for anonymization will continue to be refined and improved. In particular, incorporation of host-based sensor data is planned to supplement network logs. Additionally, there are plans to leverage Mitre's Caldera infrastructure to deploy more sophisticated, subtle attacks~\cite{mitre:caldera}.

\section*{Acknowledgments}

The authors would like to thank the late Dr. Malathi Veeraraghavan for her significant contributions to this work.